\documentclass[prd,aps,superscriptaddress,nofootinbib,notitlepage,twocolumn]{revtex4-1}

\usepackage{amsmath,graphicx,verbatim,epsfig}
\usepackage{epstopdf}
\usepackage[utf8x]{inputenc}
\usepackage[colorlinks=true,linkcolor=blue,citecolor=blue]{hyperref}

\newcommand{\nn}{\nonumber}

\newcommand{\bn}{{\bar n}}

\newcommand{\pslash}{{\not \!p}}
\newcommand{\kslash}{{\not \!k}}

\newcommand{\bnslash}{{\not \!\bn}}

\newcommand{\veuv}{\varepsilon_{\rm {UV}}}

\newcommand{\be}{\begin{equation}}
\newcommand{\ee}{\end{equation}}
\newcommand{\bea}{\begin{eqnarray}}
\newcommand{\eea}{\end{eqnarray}}
\newcommand{\balign}{\begin{align}}
\newcommand{\ealign}{\end{align}}
\newcommand{\as}{\alpha_s}

\newcommand{\sandwich}[3]{\left< #1 \right | #2 \left | #3 \right >}

\newcommand{\bg}{\begin{gather}}
\newcommand{\foma}{\end{gather}}

\newcommand{\noopsort}[1]{}

\def\L{\Lambda}

\def\z{\zeta}

\def\<{\langle}
\def\>{\rangle}

\def\a{\alpha}

\def\g{\gamma}  
\def\d{\delta}  \def\D{\Delta}
\def\l{\lambda}   \def\L{\Lambda}
\def\s{\sigma}

\def\m{\mu}
\def\n{\nu}

\def\z{\zeta}

\def\({\left(}
\def\[{\left[}
\def\){\right)}
\def\]{\right]}

\def\ln{\hbox{ln}}

\def\Qslash{Q\!\!\!\!\slash}

\def\bnslash{\bar n\!\!\!\slash}
\def\pslash{p\!\!\!\slash}
\def\bpslash{\bar p\!\!\!\slash}

\def\kslash{k\!\!\!\slash}

\def \le { \left    }
\def \ri { \right }

\def\bp{\bar p}

\def\lqcd{\L_{\rm QCD}}

\begin{document}


\title{Soft and Collinear Factorization and Transverse Momentum Dependent Parton Distribution Functions}


\author{Miguel G. Echevarr\'ia}
\email{miguel.gechevarria@fis.ucm.es}
\affiliation{Departamento de F\'isica Te\'orica II,
Universidad Complutense de Madrid (UCM),
28040 Madrid, Spain}

\author{Ahmad Idilbi}
\email{idilbi@ectstar.eu}
\affiliation{European Centre for Theoretical Studies in Nuclear Physics and elated Areas (ECT*),
Villa Tambosi, Strada delle Tabarelle 286, I-38123, Villazzano, Trento, Italy}

\author{Ignazio Scimemi}
\email{ignazios@fis.ucm.es}
\affiliation{Departamento de F\'isica Te\'orica II,
Universidad Complutense de Madrid (UCM),
28040 Madrid, Spain}




\begin{abstract}

In this work we consider how a parton distribution function, with an explicit transverse momentum dependence can be properly defined in a regularization-scheme independent manner.
We argue that by considering a factorized form of the transverse momentum dependent spectrum for the production of a heavy lepton pair in Drell-Yan reaction, one should first split the relevant soft function into two boost invariant contributions.
When those soft contributions are added to the pure collinear contributions, well-defined hadronic matrix elements emerge, i.e., the transverse momentum dependent distributions.
We also perform a comparison with Collins' definition.

\end{abstract}


\maketitle


\section{Introduction}

Transverse momentum dependent (TMD) factorization theorems are essential theoretical tools in modern high-energy colliders. 
Differential cross sections with respect to final state transverse momenta in a typical high energy processes like semi-inclusive deep inelastic scattering (SIDIS), Drell-Yan heavy lepton pair production (DY) or jet production in $p+p$ collisions can only be reliably calculated when TMD factorization theorems are established.
Such theorems have mainly dual purpose.
First is to disentangle long and short distance physics while keeping the TMD manifest. This serves to consider $q_T$-dependent spectrums in which the product(s) of high-energy reactions are produced (or tagged) with a non-vanishing transverse momentum. 
Those events constitute a major bulk of the LHC data.
Secondly, when the spin is invoked with polarized targets and/or products then the nucleon's three-dimensional structure as well as its momentum and spin distributions can be unraveled to a maximal extent.

Establishing TMD factorization theorems is much more complex than the collinear counterparts.
The complexity arises since, in certain kinematical regimes, the relevant modes have the same invariant mass but differ in their relative rapidities. 
In such instances the soft and collinear modes can mix under boosts, thus the scale factorization becomes a subtle issue. 
Working perturbatively in a specific reference frame --where it is kinematically established that the soft and collinear are the relevant modes-- then all quantities appearing in the factorized cross section have to be boost invariant so as the scale factorization remains intact under any arbitrary boost transformations.

TMD parton distribution functions (TMDPDFs) were introduced by Collins and Soper~\cite{Collins:1981uw,Collins:1981uk}.
Later on Ji, Ma and Yuan~\cite{Ji:2004wu} re-defined TMDPDFs by subtracting a complete soft function from the bi-local collinear quark field operator. 
Since then, several works have emerged~\cite{Collins:2011zzd,Chiu:2012ir,GarciaEchevarria:2011rb,Echevarria:2012qe,Cherednikov:2008ua,Collins:2012uy,Becher:2010tm,Chay:2012mh} with seemingly contradictory conclusions regarding the role of the soft function in the definition of what would be considered as an acceptable definition of a TMDPDF.

Below we argue that only when the soft function is partitioned --in rapidity space-- in a specific and subtle way and then combined with the collinear matrix elements, one can obtain a ``two-hemisphere''-like picture.
As a consequence, we are able to define TMDPDFs (polarized or unpolarized) which are free from all rapidity divergences (RDs).
Among other things, those divergences pose a serious problem to the proof of renormalizability of such physical objects.
Our approach is based on the conviction that such divergences should not exist, right from the start (and before renormalization) in the (partonic) perturbative calculation of a properly defined TMDPDF.
In this sense, we insist on treating the TMDPDF on an equal footing as the integrated PDF itself.

In~\cite{Collins:2012uy} Collins and Rogers established an equivalence between Collins' approach (JCC)~\cite{Collins:2011zzd} and the Echevarria-Idilbi-Scimemi (EIS) approach~\cite{GarciaEchevarria:2011rb}.
Below we review the latter and discuss the constraints under which the TMDPDF was defined in~\cite{GarciaEchevarria:2011rb}.  
Those constraints, which basically relate two independent collinear sectors by using the \emph{same} value of the $\D$-regulator --which will be introduced below-- limit the scope of applicability of perturbative calculations that support the ``proper'' definition of hadronic matrix elements, and prevent them from being the most general ones.  
In principle, two decoupled sectors should be treated independently which, in perturbation theory, translates into the use of different sets of regulators (with different transformation properties under boost transformations). 
Thus the definition of the TMDPDF introduced in~\cite{GarciaEchevarria:2011rb} should be generalized to the general case where no assumptions on the two decoupled collinear sectors are imposed~\footnote{This is a common practice in SCET literature, see e.g.~\cite{Manohar:2003vb,Chiu:2009yx,Chay:2012mh}}. 
We show below that this task is not a trivial one. 
Then, we establish an equivalence between this ``modified EIS'' and JCC definitions.


\section{Outline of the Problem}
\label{sec:problem}
Let us consider DY kinematics in the center-of-mass frame where the momenta of the two incoming partons initiating the hard reaction are $p$ and $\bp$. A general vector $v$ is decomposed as: $v^\m =\bn\cdot v \frac{n^\m}{2} + n\cdot v \frac{\bn^\m}{2} + v_\perp = v^+ \frac{n^\m}{2} + v^- \frac{\bn^\m}{2} + v_\perp$, where
$n = (1,0,0,1)$, $\bn = (1,0,0,-1)$.
Thus at tree level we have: $p=(Q,0,\vec 0_\perp)$ and $\bp=(0,Q,\vec 0_\perp)$ where $Q$ is the virtual photon mass.
The relevant modes that contribute to the DY $q_T$-spectrum are collinear ($k\sim Q(1,\l^2,\l)$), anti-collinear ($k\sim Q(\l^2,1,\l)$) and soft ($k\sim Q(\l,\l,\l)$), where $\l\sim q_T/Q$ is small.

As mentioned earlier the soft modes can become $n,\bn$-collinear under boosts and vice versa.
However one can still define a $n,\bn$-collinear and soft contributions which are boost invariant, as we show below.
Definitely, there is a need to introduce rapidity cuts, which also serve as regulators for rapidity divergences occurring when $y\equiv\frac{1}{2}\ln|k^+/k^-|\rightarrow \pm\infty$.

\begin{figure}[t]
\begin{center}
\includegraphics[width=0.3\textwidth]{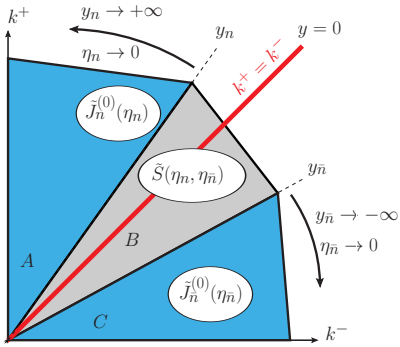}
\end{center}
\caption{\it
Relevant kinematical regions for the factorization of DY $q_T$-spectrum. Regions A and C represent the \emph{pure} collinear modes in the $n$ and $\bn$ directions, respectively, and region B represents the soft modes. $\eta_{n(\bn)}$ stand for generic rapidity regulators (and not explicit rapidity cutoffs) necessary to separate the soft and collinear modes, and at the same time, serve for regulating rapidity divergences.
When $\eta_{n(\bn)}\to 0$ we have $y_n \to +\infty$ and $y_\bn \to -\infty$.
The line $k^+=k^-$ corresponds to rapidity $y=0$.
}
\label{fig:fig}
\end{figure}

In Fig.~\ref{fig:fig} $\eta_{n(\bn)}$ are generic rapidity regulators that separate soft modes from the $n,\bn$-collinear ones.
They also serve as regulators for the rapidity divergences that appear in the collinear and soft matrix elements.
Those regulators should disappear when one combines all matrix elements within the factorization theorem, since in full QCD there are no rapidity divergences.

In terms of these generic rapidity regulators $\eta_{n(\bn)}$, the hadronic tensor for the $q_T$-dependent DY spectrum can be split in impact parameter space (IPS) as~\footnote{The tilde refers to quantities evaluated in impact parameter space and the superscript $(0)$ stands for \emph{pure} collinear contributions.}~\cite{GarciaEchevarria:2011rb}
\begin{align}\label{eq:factmodes}
\tilde M &=
H(Q^2)\,
\tilde J^{(0)}_n(\eta_n)\,
\tilde S(\eta_n,\eta_\bn)\,
\tilde J_{\bn}^{(0)}(\eta_\bn)
\end{align}
where we show explicitly just the rapidity regulator dependence. $\tilde S$ is the relevant soft function and $\tilde J_{n(\bn)}^{(0)}$ stand for \emph{pure} (anti-)collinear contributions, which are calculated first by integrating over all momentum space and then subtracting the ``zero-bin'' contribution, i.e., the soft limit of the collinear integrands~\cite{Manohar:2006nz} (see also~\cite{Chiu:2009yx}).
Generally speaking, this should be done on a diagram-by-diagram basis.
At operator level one can identify those soft contaminations with the soft function itself~\cite{Lee:2006nr,Idilbi:2007ff,Idilbi:2007yi}, however this equivalence might get spoiled for certain regulators~\footnote{Explicit examples can be found in eq.~(64) of~\cite{Manohar:2006nz}, and also in~\cite{Chiu:2012ir} where the zero-bin contributions vanish beyond tree-level while the soft function does not.}.
Thus at the level of the factorization theorem itself, one should refrain from subtracting the soft function (since this would be based on a regulator-dependent arguments) and formulate the relevant theorem in terms of pure collinear matrix elements and soft functions as in eq.~(\ref{eq:factmodes}).

At the operator level one defines the  collinear matrix element as
\begin{align}
J_n(0^+,y^-,\vec y_\perp) &= \frac{1}{2}\sum_{\s_1}
\sandwich{P\,\s_1}{\overline\Psi_{n}(0^+,y^-,\vec y_\perp)\,\frac{\bnslash}{2}\, \Psi_{n}(0)}{P\,\s_1} \,,
\end{align}
where $\Psi_{n} = W_n^\dagger T_n^\dagger \xi_n$.
$W_n$ and $T_n$ are the collinear and transverse Wilson lines, respectively, which guarantee the gauge invariance of the matrix element both in regular and singular gauges~\cite{GarciaEchevarria:2011md,Idilbi:2010im}.
In the above expression it is implied that the soft contamination has not been subtracted yet (thus $\tilde{J}^{(0)}_n\neq \tilde{J}_n$)
so we refer to it below as the ``naive'' collinear matrix element.
The relevant soft function for the $q_T$-dependent DY spectrum is given by
\begin{align}
\label{eq:soft}
S(0^+,0^-,\vec y_\perp) =
\sandwich{0}{ {\rm Tr} \; \Big[S_n^{T\dagger} S^T_\bn \Big](0^+,0^-,\vec y_\perp)\le[S^{T\dagger}_\bn S^T_n\ri](0)}{0}\,,
\end{align}
where $S_{n}(x)= P \exp \left[i g \int_{-\infty}^0 ds\, n \cdot A_s (x+s n)\right]$ is a soft Wilson line and the superscript $T$ stands for transverse Wilson lines.
The appropriate definitions of the collinear, soft and transverse Wilson lines for DY and DIS kinematics can be found in~\cite{GarciaEchevarria:2011rb}.

In~\cite{GarciaEchevarria:2011rb} the $\D$-regulator was implemented. 
We write the poles of the fermion propagators with a real and positive parameters $\D^\pm$,
\begin{align}\label{fermionsDelta}
\frac{i(\pslash+\kslash)}{(p+k)^2+i0} &\longrightarrow
\frac{i(\pslash+\kslash)}{(p+k)^2+i\D^-}\,,
\nn\\
\frac{i(\bpslash+\kslash)}{(\bp+k)^2+i0} &\longrightarrow
\frac{i(\bpslash+\kslash)}{(\bp+k)^2+i\D^+}\,,
\end{align}
and for collinear and soft Wilson lines one has
\begin{align}\label{eq:deltas}
\frac{1}{k^+\pm i0} \longrightarrow
\frac{1}{k^+\pm i\d^+}\,,
\quad\quad
\frac{1}{k^-\pm i0} \longrightarrow
\frac{1}{k^-\pm i\d^-}
\,.
\end{align}
Given the fact that the soft and collinear matrix elements should reproduce the soft and collinear limits of full QCD, then they need to be regulated consistently, so $\d^\pm$ are related with $\D^\pm$ through the large components of the collinear fields,
\begin{align} \label{regul_DeltaDY}
\d^+ = \frac{\D^+}{\bp^-}\,, \quad &\quad \quad \d^- = \frac{\D^-}{p^+} \,.
\end{align}
In particular, the soft function which, at operator level, does not know about $\d$'s (or $\D$'s), will have a dependence on $p^+$ and $\bp^-$ because in perturbation theory those regulators will be invoked. 
From the above relations and by considering the denominators of the propagators, it is clear that one should treat the $\Delta^{\pm}$ as boost invariant quantities, i.e., they transform as the product $p^+ \bar{p}^-$, while $\delta^{+}$ transforms as $k^+$ or $p^+$ (or $1/\bar{p}^-$) and $\delta^-$ transforms as $k^-$ or $\bar {p}^-$ (or $1/p^+$). Those observations will be used below. 

Using the $\D$-regulator one can relate $\eta_n$ with $\D^-$ and $\eta_\bn$ with $\D^+$ \emph{only} in the terms where the $\D$'s regularize rapidity divergences, but not in the terms with IR divergences (which are also regularized by $\D^{\pm}$).
When all matrix elements are combined in eq.~(\ref{eq:factmodes}) there will remain a $\D$-dependence which is exactly the genuine IR divergences of perturbative QCD. This remaining $\D$-dependence is not worrisome since, hadronically, it disappears by confinement, i.e., due to non-perturbative QCD contributions in exactly the same manner as the collinear divergence of the partonic integrated PDF signals the onset of non-perturbative (long distance) contribution.
The distinction between rapidity divergences and the IR ones will become more clear in the following section, where we show explicit results for the collinear and soft matrix elements and comment on them.

In~\cite{GarciaEchevarria:2011rb} the soft function appearing in eq.~(\ref{eq:factmodes}) was split identically between the two collinear sectors and the TMDPDFs were defined as
\begin{align}\label{eq:tmddefold}
\tilde F_n^{\rm old} &=
\le.\tilde J^{(0)}_n(\D^-)\sqrt{\tilde S\left(\frac{\D^-}{p^+},\frac{\D^+}{\bar {p}^-}\right)}\ri |_{\D^+=\D^-}
\,,
\nn\\
\tilde F_\bn^{\rm old} &=
\le.\tilde J^{(0)}_\bn(\D^+)\sqrt{\tilde S\left(\frac{\D^-}{p^+},\frac{\D^+}{\bar {p}^-}\right)}\ri |_{\D^+=\D^-}
\,.
\end{align}
where the $\D$-regulator was used to regularize all the IR and the rapidity divergences.
The ${\cal O}(\as)$ result for the TMDPDF, before renormalization, is
\begin{multline}\label{eq:tmdold2}
\tilde F_{n1}^{\rm old} =
\frac{\alpha_s C_F}{2\pi}\le\{
\d(1-x_n)\le[
\frac{1}{\veuv^2}
+ \frac{3}{2\veuv}- \frac{1}{\veuv}\ln\frac{Q^2\D^-}{\m^2\D^+}
\ri.\ri.
\\
\le.\le.
- \frac{1}{2}L_\perp^2 + \frac{3}{2}L_\perp
- L_\perp\ln\frac{Q^2\D^-}{\m^2\D^+} - \frac{\pi^2}{12}
\ri]
+ (1-x_n) - L_\perp {\cal P}_{q/q}
\ri.
\\
\le.
- {\cal P}_{q/q} \ln\frac{\D^-}{\m^2} - \frac{1}{4}\d(1-x_n) - (1-x_n)[1+\ln(1-x_n)] \ri\}
\,.
\end{multline}
where $L_\perp=\ln(\m^2b^2e^{2\g_E}/4)$, ${\cal P}_{qq}$ is the Altarelli-Parisi quark-quark splitting kernel and $\D^+=\D^-$ as shown in eq.~(\ref{eq:tmddefold}).
We label this definition as ``old'' as compared to the one given in this paper.
The new definition will be a generalization of this one in the sense that no assumption will be made on the values of $\D^\pm$, although it reduces to the one in eq.~(\ref{eq:tmddefold}) for $\D^+=\D^-$.

It is important to emphasize that in principle two disentangled collinear sectors should be treated independently, thus it is crucial to examine what happens when we relax the condition $\D^+=\D^-$  and consider the most general case where $\D^+\neq \D^-$.

The first line in eq.~(\ref{eq:tmdold2}) contains a mixed UV-RD term. 
The second line is what would be the matching (or Wilson) coefficient of the TMDPDF onto the integrated PDF after an OPE is carried out, but it also contains an unacceptable dependence on the $\D$-regulator, i.e., an un-cancelled RD.
As is well-known, Wilson coefficients should be free from any non-ultraviolet regulators, either IR or RD.
The last line is simply the integrated PDF (see ~eq.~(A.6) in~\cite{GarciaEchevarria:2011rb}).
In the next section we comment on the origin of the $Q^2$-dependence appearing above.

If we combine eq.~(\ref{eq:tmdold2}) with the analogous result for $\tilde{F}_{\bn 1}^{\rm old}$ (where we just interchange $\D^+ \leftrightarrow \D^-$ and $x_n \leftrightarrow x_\bn$) the RDs cancel and the hard part $H$ in $\tilde{M}=H\,\tilde{F}_n^{\rm old}\,\tilde{F}_\bn^{\rm old}$ depends just on $Q^2/\m^2$, as it should.
However it is clear that only with the choice $\D^+=\D^-$ the RDs cancel in each TMDPDF independently, both the mixed UV-RD divergence in the first line and the RD in the Wilson coefficient in the second line.
It is important to notice that the limit
\begin{align}\label{eq:limit}
\lim_{\substack{\D^+\to 0 \\ \D^- \to 0}} \ln\frac{\D^-}{\D^+}\,,
\end{align}
has to be taken in order to get a well-defined physical quantity.
Apart from the $\D^-$-dependence in the last line of eq.~(\ref{eq:tmdold2}), which is the manifestation of the genuine long-distance QCD effects and is washed out by confinement, all the remaining $\ln(\frac{\D^-}{\D^+})$ in the first two lines should cancel in that limit.
But clearly this is not the case.
The independent behavior of $\D^+$ and $\D^-$, which is part  of the implementation of the $\D$-regulator (reminiscent from taking the full QCD propagators with $\D^\pm$ into the soft and collinear limits) renders that limit as ill-defined since it could either be finite or $\pm\infty$.
From the above it is thus clear that in order to avoid any ad-hoc prescription for the regulators ($\D^+=\D^-$) a new definition of the TMDPDF should be adopted.
The new definition should nonetheless reduce to the one discussed above when $\D^+=\D^-$.
Finally we point out that the discussion of the equivalence between JCC and EIS approaches in~\cite{Collins:2012uy} is valid only when the limit in eq.~\ref{eq:limit} is finite.
In this paper we generalize the arguments in~\cite{Collins:2012uy} to the general case where there is no relation between $\D^+$ and $\D^-$ and $\lim_{\substack{\D^\pm\to 0 }} |\ln(\D^-/\D^+)|=\infty$.

Although our presentation so far was done in terms of the $\D$-regulator, our results, to be presented below, can be immediately generalized to other regulators as well.
If one had used off-shellnesses~\cite{Chay:2012mh}, or the $\nu$-regulator as in~\cite{Chiu:2012ir}, then our proposed TMDPDF would have the same features as with the $\D$-regulator.
This has been checked explicitly.
The implementation of the analytic regulator used in~\cite{Becher:2010tm} needs further investigation and will be studied elsewhere.
We will show next that by splitting the soft function in two ``pieces'' (and not taking naively its square root), which will turn out to be a fundamental property of it that holds to all orders in perturbation theory, and by combining them with the collinear matrix elements, we will be able to properly define the TMDPDF and cancel rapidity divergences.

\section{Splitting the Soft Function and Definition of TMDPDF}
\label{sec:splitting}
In the kinematical region where $Q\gg q_T\gg\lqcd$ one  can perform an operator product expansion (OPE) of the result in eq.~(\ref{eq:factmodes}) onto the integrated PDFs where the hadronic tensor can be expressed as~\footnote{$\tilde C(x_n,x_{\bn};L_\perp,Q^2/\m^2)=\tilde C_n^{\Qslash}(x_n;L_\perp)\,
\tilde C_\bn^{\Qslash}(x_\bn;L_\perp)\,
\le(\frac{Q^2}{\m^2}\ri)^{-2D\le(\as(\m),L_\perp(\m)\ri)}$ in the notation of~\cite{GarciaEchevarria:2011rb}.}
\begin{align}\label{eq:prove1}
\tilde M &=
H(Q^2/\m^2)\,
\tilde C(x_n,x_{\bn};L_\perp,Q^2/\m^2)
\nn\\
&\times
f_n(x_n;\D^-/\m^2)\,
f_{\bn}(x_\bn;\D^+/\m^2)
\,.
\end{align}
The functions $H$ and $\tilde C$ are the two perturbatively calculable matching coefficients obtained after a two-step matching at the scales $Q$ and $q_{\rm T}$, respectively.
Those coefficients are independent of any non-ultraviolet regulators.
In particular, using the $\D$-regulator, they are independent of the $\D^{\pm}$.

Since the integrated PDFs, $f_{n(\bn)}$, contain just the $n,\bn$-IR collinear divergences then each one of them can be written in general as
\begin{align}\label{eq:pdflog}
\ln f_{n} &= {\cal R}_{f1}(x_n,\as) + {\cal R}_{f2}(x_n,\as)\ln\frac{\D^-}{\m^2}
\,,
\nn\\
\ln f_{\bn} &= {\cal R}_{f1}(x_{\bn},\as) + {\cal R}_{f2}(x_\bn,\as)\ln\frac{\D^+}{\m^2}\,,
\end{align}
where ${\cal R}_{f1}$ and ${\cal R}_{f2}$ are some functions of $\as(\m)$ and $x_{n(\bn)}$.
The fact that $\ln f_n$ has only a single $\ln\frac{\D^-}{\m^2}$ (or single IR pole in pure dimensional regularization) to all orders in perturbation theory is a well-known fact.
It has been shown~\cite{Korchemsky:1987wg} (see also~\cite{Manohar:2003vb}) that the anomalous dimension of the PDF in Mellin moment space has a single logarithm  $\ln N$ to all orders in perturbation theory.
This single logarithm  results from a single UV pole in  $\ln f_n$.
For a massless matrix element such as the PDF, the single UV pole will always be accompanied by a single IR pole in dimensional regularization, or a single $\ln\frac{\D^-}{\m^2}$ if one uses the $\D$-regulator for the IR divergences.

On the other hand we can express the hadronic tensor $\tilde M$ in terms of matrix elements, as in~eq.~(\ref{eq:factmodes}).
To separate the modes in rapidity we notice that each one of the pure collinear matrix elements does not have any information about the other collinear sector.
This is exactly due to the fact that the soft contamination in the naively calculated collinear contribution has been subtracted out.
By its definition, it is clear that the soft function depends on both sectors, which is manifested through the dependence on both $\D$'s.
Given this, and using boost invariance and dimensional analysis we can write
\begin{align}\label{eq:prove2}
\ln\tilde{J}^{(0)}_{n} &=
{\cal R}_n\le(x_n,\as,L_\perp,\ln\frac{\D^-}{\m^2}\ri)
\,,
\nn\\
\ln\tilde{J}^{(0)}_{\bn} &=
{\cal R}_{\bn}\le(x_\bn,\as,L_\perp,\ln\frac{\D^+}{\m^2}\ri)
\,,
\nn\\
\ln\tilde{S} &=
{\cal R}_s\le(\as,L_\perp,\ln\frac{\D^-\D^+}{Q^2\m^2}\ri)\,,
\end{align}
where ${\cal R}_{(n,\bn,s)}$ stand for generic functions.

Now, given that all the IR divergences of QCD are absorbed into two PDFs, as shown in eq.~(\ref{eq:prove1}), given the single logarithmic structure of the PDFs given eq.~(\ref{eq:pdflog}) and that the pure collinear matrix elements $\tilde{J}^{(0)}$ depend just on one of the regulators (the relevant for each sector), as shown in eq.~(\ref{eq:prove2}), combined with the symmetry between $n$ and $\bn$ in the soft function, we immediately deduce that~eq.~(\ref{eq:prove2}) can be rewritten as
\begin{align}\label{eq:prove3}
\ln\tilde{J}^{(0)}_{n} &=
{\cal R}_{n1}(x_n,\as,L_\perp) + {\cal R}_{n2}(x_n,\as,L_\perp)\,\ln\frac{\D^-}{\m^2}
\,,
\nn\\
\ln\tilde{J}^{(0)}_{\bn} &=
{\cal R}_{\bn 1}(x_\bn,\as,L_\perp) + {\cal R}_{\bn 2}(x_\bn,\as,L_\perp)\,\ln\frac{\D^+}{\m^2}
\,,
\nn\\
\ln\tilde{S} &=
{\cal R}_{s1}(\as,L_\perp) + {\cal R}_{s2}(\as,L_\perp)\,\ln\frac{\D^-\D^+}{Q^2\m^2}
\,.
\end{align}

Before we continue our discussion of the splitting of the soft function based on eq.~(\ref{eq:prove3}), let us comment, as promised before, on the $Q^2$-dependence in that function.
The arguments that led to eq.~(\ref{eq:prove3}) do not specify this dependence by themselves and an additional input is needed.
Actually and just by looking at the product of the two regulators in the soft function, $\D^+/{\bar p}^-$ and $\D^-/p^+$, one would deduce that $\tilde{S}$ is function of $\hat s \equiv (p+\bar p)^2=p^+\bar p^-$ rather than $Q^2$.
The partonic invariant mass $\hat{s}$ is related to $Q^2$ by the relation $x_n x_\bn=Q^2/\hat{s}$ where $x_n=\sqrt{Q^2/\hat{s}}\,e^y$, $x_{\bn}=\sqrt{Q^2/\hat{s}}\,e^{-y}$ and $y$ is the rapidity of the produced virtual photon or, equivalently, of the heavy lepton pair.
These relations for $x_n$ and $x_{\bn}$ are valid in the small $q_T$-limit and they have corrections~\footnote{For arbitrary $q_T$ one has $x_n=\sqrt{(Q^2+q_T^2)/\hat{s}}\,e^y$ and $x_\bn=\sqrt{(Q^2+q_T^2)/\hat{s}}\,e^{-y}$.} of order $q_T/Q$ which are of order $\lambda$ in the effective theory and thus can be neglected.
By simple kinematics one can show that the inequality between $\hat{s}$ and $Q^2$ resulting from soft gluon radiation is of order $\lambda$, or in other words, $\hat{s}=Q^2+{\cal{O}}(Q^2\lambda)$.
To leading order in $\lambda$ we can thus safely write: $\hat{s}=p^+{\bar p}^- =Q^2$ in any contribution of the soft function  and to all orders in perturbation theory.

Two immediate conclusions arise from the above analysis.
First is that the soft function can be considered, at leading order in $\l$, as function only of $Q^2$, as claimed in eq.~(\ref{eq:prove3}).
And second is that its contribution to the hadronic tensor $\tilde{M}$ in momentum space will always be accompanied by the product $\delta(1-x_n)\delta(1-x_{\bn})$ (for an explicit ${\cal O}(\as)$ calculation see, e.g., eq.~(19) in~\cite{Idilbi:2005er}).

Since the $Q^2$-dependence of the soft function has been established, we can go back to eq.~(\ref{eq:prove3}) and make the following splitting: $\ln\frac{\D^-\D^+}{Q^2\m^2}=\frac{1}{2}\ln\frac{\alpha(\D^-)^2}{Q^2\m^2}+\frac{1}{2}\ln\frac{(\D^+)^2}{\alpha Q^2\mu^2}$ and thus we are led to write the following two quantities:
\begin{multline}
\ln\tilde{S}\le(\frac{\D^-}{p^+},\a\frac{\D^-}{\bp^-}\ri) =
\ln\tilde{S}\le(\a\frac{\D^-}{p^+},\frac{\D^-}{\bp^-}\ri) =
\\
{\cal R}_{s1}(\as,L_\perp)
+ {\cal R}_{s2}(\as,L_\perp)\,\ln\frac{\a (\D^-)^2}{Q^2\m^2}
\end{multline}
and
\begin{multline}
\ln\tilde{S}\le(\frac{1}{\a}\frac{\D^+}{p^+},\frac{\D^+}{\bp^-}\ri) =
\ln\tilde{S}\le(\frac{\D^+}{p^+},\frac{1}{\a}\frac{\D^+}{\bp^-}\ri) =
\\
{\cal R}_{s1}(\as,L_\perp)
+ {\cal R}_{s2}(\as,L_\perp)\,\ln\frac{(\D^+)^2}{\a Q^2\m^2}
\,,
\end{multline}
which means that to all orders in perturbation theory the complete soft function $\tilde{S}$ can be split according to
\begin{multline}\label{eq:softsplitting_log}
\ln\tilde{S}\le(\frac{\D^-}{p^+},\frac{\D^+}{\bp^-}\ri) =
\\
\frac{1}{2}\ln\tilde{S}\le(\frac{\D^-}{p^+},\a\frac{\D^-}{\bp^-}\ri)
+ \frac{1}{2}\ln
\tilde{S}\le(\frac{1}{\a}\frac{\D^+}{p^+},\frac{\D^+}{\bp^-}\ri).
\end{multline}
Notice that the above equation holds in the limits $\D^\pm\to 0$, which are uncorrelated.
Also take into account that the arbitrariness in the splitting of the single logarithm of the soft function in eq.~(\ref{eq:prove3}) manifests itself as the parameter $\a$, which is a boost invariant real number and it is always finite (even when $\lim_{\D^\pm\rightarrow 0}|\ln\D^-/\D^+|=\infty$).
Since the soft function can indeed be separated into two ``pieces'', we define the TMDPDFs as
\begin{align}\label{eq:tmdnewdef}
\tilde F_{n}(x_n,b;\sqrt{\z_n},\m) &=
\tilde J^{(0)}_{n}(\D^-)\,
\sqrt{\tilde{S}\le(\frac{\D^-}{p^+},\a\frac{\D^-}{\bp^-}\ri)}\,,
\nn\\
\tilde F_{\bn}(x_\bn,b;\sqrt{\z_\bn},\m) &=
\tilde J^{(0)}_{\bn}(\D^+)\,
\sqrt{\tilde{S}\le(\frac{1}{\a}\frac{\D^+}{p^+},\frac{\D^+}{\bp^-}\ri)}
\,,
\end{align}
where $\z_n=Q^2/\a$ and $\z_\bn=\a Q^2$, and thus $\z_n\z_\bn = Q^4$.
This parameter $\z_{n(\bn)}$ is equivalent to the one that appears in JCC formalism~\cite{Collins:2011zzd}.
The soft function was calculated in~\cite{GarciaEchevarria:2011rb} with the $\D$-regulator, and its result at ${\cal O}(\as)$ is
\begin{multline}\label{eq:softfirst}
\tilde S_1\le(\frac{\D^-}{p^+},\frac{\D^+}{\bp^-}\ri) =
\frac{\alpha_s C_F}{2\pi} \le[
- \frac{2}{\veuv^2} + \frac{2}{\veuv} \ln\frac{\D^-\D^+}{\m^2Q^2}
\ri.
\\
\le.
+ L_\perp^2
+ 2L_\perp \ln\frac{\D^-\D^+}{\m^2Q^2} + \frac{\pi^2}{6}
\ri]
\\
= \frac{1}{2}\le[
\tilde{S}_1\le(\frac{\D^-}{p^+},\a\frac{\D^-}{\bp^-}\ri)
+ \tilde{S}_1\le(\frac{1}{\a}\frac{\D^+}{p^+},\frac{\D^+}{\bp^-}\ri)
\ri]
\,,
\end{multline}
thus establishing eq.~(\ref{eq:softsplitting_log}) at ${\cal O}(\alpha_s)$.

We next consider the ${\cal{O}}(\alpha_s)$ results for the TMDPDF, defined in eq.~(\ref{eq:tmdnewdef}), given the splitting of the soft function.

The naive collinear matrix element was calculated in~\cite{GarciaEchevarria:2011rb} and we have
\begin{multline}\label{eq:naive}
\tilde J_{n1} =
\frac{\alpha_s C_F}{2\pi} \le\{
\d(1-x_n)\le[
\frac{2}{\veuv}\ln\frac{\D^+}{Q^2} + \frac{3}{2\veuv}
\ri.\ri.
\\
\le.\le.
- \frac{1}{4} + \frac{3}{2}L_\perp + 2L_\perp\ln\frac{\D^+}{Q^2}
\ri]
\ri.
\\
\le.
- (1-x_n)\ln(1-x_n)
- {\cal P}_{qq} \ln\frac{\D^-}{\m^2}
- L_\perp {\cal P}_{qq}
\ri\}\,,
\end{multline}
where the $\D^-$ that appears in combination with the splitting function ${\cal P}_{qq}$ is pure IR, while the other $\D^+$ serve as rapidity regulators and can be identified with the generic rapidity regulator $\eta_{\bn}$ mentioned before.
This $\D^+$ dependence comes from the regularization of the collinear Wilson line $W_n$, as in eq.~(\ref{eq:deltas}) (see~\cite{GarciaEchevarria:2011rb} for more details).
One would expect that the $n$-collinear matrix element depends on the rapidity regulator that belongs to that sector, i.e., $\D^-$ (or $\eta_n$ in general).
However, due to the fact that the naive collinear contains soft contamination, it also depends on $\D^+$.
By subtracting the zero-bin~\footnote{With the $\D$-regulator it was shown in~\cite{GarciaEchevarria:2011rb} that the subtraction of the zero-bin is equivalent to divide by the soft function.} this dependence is switched back to the proper parameter, $\D^-$, and the pure collinear matrix element is given by
\begin{multline}\label{eq:pure}
\tilde J_{n1}^{(0)} =
\frac{\alpha_s C_F}{2\pi} \le\{
\d(1-x_n)\le[
\frac{2}{\veuv^2} - \frac{2}{\veuv}\ln\frac{\D^-}{\m^2} + \frac{3}{2\veuv}
\ri.\ri.
\\
\le.\le.
- \frac{1}{4} - \frac{2\pi^2}{12} - L_\perp^2 + \frac{3}{2}L_\perp - 2L_\perp\ln\frac{\D^-}{\m^2}
\ri]
\ri.
\\
\le.
- (1-x_n)\ln(1-x_n)
- {\cal P}_{qq} \ln\frac{\D^-}{\m^2}
- L_\perp {\cal P}_{qq}
\ri\}
\end{multline}
which, as explained before, depends only on $\D^-$.
Again we emphasize that the $\D^-$ accompanying the splitting function ${\cal P}_{q/q}$ is pure IR, while the other $\D^-$-dependent terms include RDs.

Combining the pure collinear $\tilde{J}_{n1}^{(0)}(\D^-)$ with $\tilde S_1(\D^-/p^+,\a\D^-/\bp^-)$ that can be extracted from eq.~(\ref{eq:softfirst}), the newly defined TMDPDF given in eq.~(\ref{eq:tmdnewdef}) is
\begin{multline}\label{eq:tmdoneloop}
\tilde F_{n1}(x_n,b;\sqrt{\z_n},\m) =
\\
\frac{\alpha_s C_F}{2\pi}\le\{
\d(1-x_n)\le[
\frac{1}{\veuv^2} - \frac{1}{\veuv}\ln\frac{\z_n}{\m^2}
+ \frac{3}{2\veuv}
\ri.\ri.
\\
\le.\le.
- \frac{1}{2}L_\perp^2 + \frac{3}{2}L_\perp
- L_\perp\ln\frac{\z_n}{\m^2} - \frac{\pi^2}{12}
\ri]
+ (1-x_n) - L_\perp {\cal P}_{qq}
\ri.
\\
\le.
- {\cal P}_{qq} \ln\frac{\D^-}{\m^2} - \frac{1}{4}\d(1-x_n) - (1-x_n)[1+\ln(1-x_n)] \ri\}
\,.
\end{multline}
As the above equation shows there are no more rapidity divergences, as promised, thus it is straightforward to renormalize the TMDPDF.
The anomalous dimension of a TMDPDF acquires an explicit $Q^2$-dependence, contrary to the integrated PDF.
Our perturbative calculation for the TMDPDF for the general case where $\D^+ \neq \D^-$ indicates explicitly that the TMDPDF is boost invariant.
Moreover, it is worthwhile mentioning the disappearance of the $\D$-dependence from the matching coefficient of the TMDPDF onto the integrated PDF (the second line in the previous equation). 

At this stage it is worth mentioning that although the TMDPDF definition, eq.~(\ref{eq:tmdnewdef}), is given with the $\D$-regulator, it can be straightforwardly expressed when other commonly used regulators are implemented to regularize divergences (other than the UV ones). 
This can be established by considering the regulators for the two independent collinear sectors, their mass dimensions and their transformation properties under boosts.

With the above  definitions of TMDPDFs, the hadronic tensor for the $q_T$-dependent spectrum of DY heavy lepton-pair production at $q_T\ll Q$ can be expressed in terms of a hard part and two TMDPDFs and without a soft function
\begin{multline}\label{eq:factth}
\tilde{M}(x_n,x_\bn,b;Q^2) =
\\
H(Q^2/\m^2)\,
\tilde{F}_n(x_n,b;\sqrt{\z_n},\m)\,
\tilde{F}_\bn(x_\bn,b;\sqrt{\z_\bn},\m)
\\
+ {\cal O}\le((bQ)^{-1}\ri)
\,.
\end{multline}

\section{Equivalence of EIS and JCC Definitions of TMDPDF}
\label{sec:equivalence}

In this section we establish the equivalence between Collins' definition of TMDPDF~\cite{Collins:2011zzd} and ours given in eq.~(\ref{eq:tmdnewdef}).
In~\cite{Collins:2012uy} Collins and Rogers have already discussed this equivalence, however they considered the definition given in eq.~(\ref{eq:tmddefold}) assuming that $\lim_{\substack{\D^\pm\to 0 }} \ln(\D^-/\D^+)$ was finite.
In the following, by using the splitting of the soft function given in eq.~(\ref{eq:softsplitting_log}), we show that this equivalence also holds in the most general case where the two regulators are completely independent.
In this case, $\lim_{\substack{\D^\pm\to 0 }} |\ln(\D^-/\D^+)|$ can be also $\infty$.

The definition of the TMDPDF given in~\cite{Collins:2011zzd} is
\begin{align}\label{eq:tmddefjcc}
\tilde{F}_n^{\rm JCC}(x_n,b;\sqrt{\z_n},\m) &=
\lim_{\substack{y_n \to +\infty \\ y_\bn \to -\infty}}
\tilde{J}_n(y_\bn)\,
\sqrt{\frac{\tilde{S}(y_n,y_c)}{\tilde{S}(y_c,y_\bn)\,\tilde{S}(y_n,y_\bn)}}
\,,
\end{align}
where $\z_n = (p^+)^2 e^{-2y_c}$ and the soft functions depend on the boost invariant rapidity difference of their respective arguments, i.e., $\tilde{S}(y_1,y_2)=\tilde{S}(y_1-y_2)$.
In this definition it is assumed that the subtraction of the zero-bin contribution is equivalent to divide the naive collinear matrix element by the soft function. In the work of Collins this is justified~\cite{Collins:2011zzd} since no regulators are implemented other than rapidity cuts.
However, as already mentioned before, this is not the general case.

\begin{figure}[t]
\begin{center}
\includegraphics[width=0.3\textwidth]{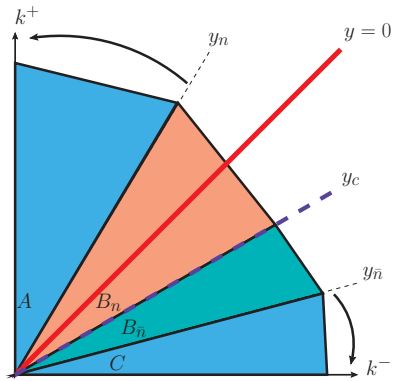}
\end{center}
\caption{\it
Rapidity regions for JCC definition of TMDPDF in eq.~(\ref{eq:tmddefjcc}).
$B_n+B_\bn$ regions represent the complete soft function $\tilde{S}(y_n,y_\bn)$.
The naive collinear $\tilde{J}_n(y_\bn)$ is represented by regions $A+B_n+B_\bn$.
Analogously, $\tilde{J}_{\bn}$ by regions $B_n+B_\bn+C$.
}
\label{fig:regions5_jcc}
\end{figure}

Looking at Fig.~\ref{fig:regions5_jcc} we can easily understand the origin of each factor in the above definition.
The naive collinear matrix element $\tilde{J}_n(y_\bn)$ is represented pictorially by regions $A+B_n+B_\bn$, which contain modes with rapidities between $+\infty$ and $y_\bn$.
$\tilde{S}(y_n,y_c)$ is represented by region $B_n$ and contains modes with rapidities between $y_n$ and $y_c$.
Similarly, $\tilde{S}(y_c,y_\bn)$ is represented by region $B_\bn$, and finally the complete soft function $\tilde{S}(y_n,y_\bn)$ is the combination of regions $B_n+B_\bn$ containing modes with rapidities between $y_n$ and $y_\bn$.
Joining all the ``pieces'' together we see that, basically, the TMDPDF $\tilde{F}_n^{\rm JCC}$ is defined as the quantity which contains the modes with rapidities between $+\infty$ and $y_c$, i.e., regions $A+B_n$.
Therefore, the other TMDPDF $\tilde{F}_\bn^{\rm JCC}$ will contain modes with rapidities between $y_c$ and $-\infty$.

Thus, based on the discussion above, one could naively think of defining the TMDPDF directly as
\begin{align}\label{eq:tmddefjccnaive}
\tilde{F}_n^{\rm JCC(naive)}(x_n,b;\sqrt{\z_n},\m) &=
\lim_{\substack{y_\bn \to -\infty}}
\frac{\tilde{J}_n(y_\bn)}{\tilde{S}(y_c,y_\bn)}
\,.
\end{align}
However, although this quantity contains modes with rapidities between $+\infty$ and $y_c$ (regions $A+B_n$), it suffers from un-cancelled self-energies at finite $y_c$.
In fact, the only purpose of the cumbersome combination of the 3 soft functions in eq.~(\ref{eq:tmddefjcc}) is to cancel the self-energy at $y_c$, but apart from this issue, the goal of the whole square root factor is simply to subtract $\tilde{S}(y_c,y_\bn)$ from the naive collinear, i.e., region $B_\bn$ in Fig.~\ref{fig:regions5_jcc}.
Notice that when one insists on keeping all Wilson lines on-the-light-cone the issue of self-energies becomes irrelevant since all self-energies cancel due to $n^2=\bn^2=0$.
Definitely, however, one needs then to introduce a set of regulators to regularize all the non-ultraviolet divergences, i.e., IR and RD.
Actually, the introduction of such regulators in perturbative calculations is a must, at least in order to carry out perturbative calculations beyond ${\cal{O}}(\alpha_s)$, where relying on cancellation of rapidity divergences between the naive collinear and a soft function just by combining integrands (see p. 389 in~\cite{Collins:2011zzd}) becomes almost an impossible task.
Moreover it also simplifies the soft factor needed to properly define a TMDPDF, as it is clear from eq.~(\ref{eq:tmdnewdef}).
The aim of the combination of collinear and soft matrix elements in eq.~(\ref{eq:tmddefjcc}) is the cancellation of the rapidity divergence when $y_\bn\to -\infty$, as eq.~(\ref{eq:tmddefjccnaive}) suggests, and the introduction of more soft factors in the definition does not introduce a rapidity divergence when $y_n\to +\infty$, since this is cancelled under the square root.

It was argued in~\cite{Collins:2012uy} the equivalence between JCC and EIS definitions of the TMDPDF by considering the definition given in eq.~(\ref{eq:tmddefold}).
That equation can be written in terms of the naive collinear matrix element, making more clear the comparison with JCC definition,
\begin{align}\label{eq:tmdolddef2}
\tilde F_n^{\rm old} &=
\frac{\tilde{J}_n(\D^+)}
{\sqrt{\tilde S\left(\frac{\D^-}{p^+},\frac{\D^+}{\bar {p}^-}\right)}}
\,.
\end{align}
Notice that we have written explicitly the dependence of $\tilde{J}_n$ on its rapidity regulator $\D^+$, but as shown in eq.~(\ref{eq:naive}), it also contains a pure IR dependence on $\D^-$.
Although a different regularization method is used, i.e., the $\D$-regulator, the naive collinear matrix element $\tilde{J}_n(\D^+)$ again is represented by regions $A+B_n+B_\bn$ in Fig.~\ref{fig:regions5_jcc}, containing the modes with rapidities between $+\infty$ and $y_\bn$.
From the result in eq.~(\ref{eq:softfirst}) for the soft function and the fact that it is boost invariant, we deduce that it depends on the boost invariant rapidity difference
\begin{align}
y_n - y_\bn
=
\ln\frac{\m^2Q^2}{\D^-\D^+}
\,,
\end{align}
where the rapidity cutoffs are
\begin{align}\label{eq:ynybn}
y_n &= \ln\frac{\m p^+}{\D^-}
\,,
\quad\quad\quad
y_\bn = \ln\frac{\D^+}{\m \bp^-}
\,.
\end{align}
When taking the limits $\D^\pm\to 0$ the two rapidities $y_n$ and $y_\bn$ take also their proper limits, $y_n \rightarrow +\infty$ and $y_{\bn}\rightarrow -\infty$.
In terms of these cutoffs, eq.~(\ref{eq:tmdolddef2}) can be rewritten as
\begin{align}\label{eq:tmdolddef3}
\tilde F_n^{\rm old} &=
\frac{\tilde{J}_n(y_\bn)}
{\sqrt{\tilde S\left(y_n,y_\bn\right)}}
\,,
\end{align}
which can be more easily compared to JCC definition in eq.~(\ref{eq:tmddefjcc}).
The authors in~\cite{Collins:2012uy} showed that this two definitions are equivalent if the limits of $y_n$ and $y_\bn$ are coordinated in such a way that $y_c=(y_n+y_\bn)/2$ is finite.
In terms of the $\D$-regulator this translates into the coordination of the limits of $\D^+$ and $\D^-$, i.e.,
\begin{align}
y_c &=
\lim_{\substack{y_n\to +\infty \\ y_\bn \to -\infty}}
\frac{1}{2}(y_n+y_\bn)
=
\lim_{\substack{\D^-\to 0 \\ \D^+ \to 0}}
\frac{1}{2} \le( \ln\frac{\m\,p^+}{\D^-} + \ln\frac{\D^+}{\m\,\bp^-} \ri)
=
\nn\\
&
\lim_{\substack{\D^-\to 0 \\ \D^+ \to 0}}
\frac{1}{2}\ln\frac{\D^+}{\D^-}\frac{p^+}{\bp^-}
\,.
\end{align}

\begin{figure}[t]
\begin{center}
\includegraphics[width=0.3\textwidth]{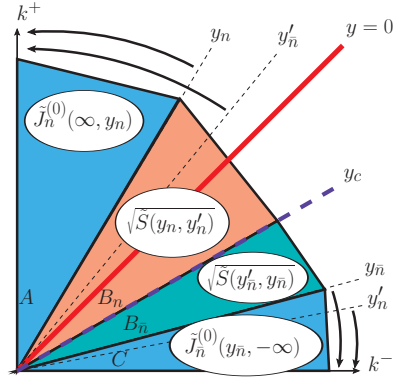}
\end{center}
\caption{\it
The splitting of the soft function at rapidity $y_c$, which is unambiguously defined by using the auxiliary lines $y'_n$ and $y'_\bn$.
}
\label{fig:regions5}
\end{figure}

However, in the general case where there is no relation between the collinear sectors, the ratio $\D^+/\D^-$ is ill-defined, and one has to resort to the splitting of the soft function shown before to properly define the TMDPDFs without making any assumption about the regulators.
In this way, we generalize the equivalence between JCC and EIS definitions shown in~\cite{Collins:2012uy}.
The splitting of the soft function given in eq.~(\ref{eq:softsplitting_log}) can be rewritten as
\begin{multline}\label{eq:softsplitting2}
\ln\tilde{S}\le(\frac{\D^-}{p^+},\frac{\D^+}{\bp^-}\ri) =
\\
\frac{1}{2}\ln\tilde{S}\le(\frac{\D^-}{p^+},\n\frac{\D^-}{p^+}\ri)
+\frac{1}{2}\ln\tilde{S}\le(\frac{1}{\n}\frac{\D^+}{\bp^-},\frac{\D^+}{\bp^-}\ri)
\,
\end{multline}
where in this case $\n=\a (p^+/\bp^-)$ is a finite and dimensionless parameter which transforms as $(p^+)^2$ under boosts, contrary to the already defined $\a$, which is a boost invariant real number.
Defining
\begin{align}
y'_n &= \ln\frac{\n\D^-}{\m p^+}
\,,
\quad\quad\quad
y'_\bn = \ln\frac{\n\m\bp^-}{\D^+}
\,,
\end{align}
and using $y_n$ and $y_\bn$ given in eq.~(\ref{eq:ynybn}), we can now rewrite
\begin{align}
y_c &=
\lim_{\substack{y_n\to +\infty \\ y'_n \to -\infty}}
\frac{1}{2}(y_n+y'_n)
=
\lim_{\substack{y'_\bn\to +\infty \\ y_\bn \to -\infty}}
\frac{1}{2}(y'_\bn+y_\bn)
=
\frac{1}{2}\ln\n
\,,
\end{align}
which is a well-defined and finite rapidity in the limit $\D^\pm \to 0$, without imposing any relation between $\D^+$ and $\D^-$.
Thus, $\z_n=(p^+)^2e^{-2y_c}$ and $\z_\bn=(\bp^-)^2e^{2y_c}$, as they appear in JCC approach.
As it is shown pictorially in Fig.~\ref{fig:regions5}, the limits of $y_n$ and $y_n'$ on one hand, and $y_\bn$ and $y_\bn'$ on the other, are coordinated and thus one can calculate their mean $y_c$.
In terms of the $\D$-regulator, $y_n$ ($y_\bn$) and $y_n'$ ($y_\bn'$) both involve the same parameter $\D^-$ ($\D^+$), and then their limits are not independent.

To conclude, we have shown that the JCC and EIS definitions of the TMDPDF lead to a properly (and equivalent) defined TMDPDF. 
The fundamental fact, shared in both approaches, is the need to include a soft function contribution to the naive collinear matrix elements while taking into account the issue of double counting among soft and collinear modes through ``soft subtraction''.
Although Collins approach is conceptually accurate, it is extremely difficult to be implemented in perturbation theory beyond one-loop due to the lack of introduction of regulators in the collinear and soft sectors. 
As explained above, when such regulators are introduced, then one needs to split the complete soft function in a subtle way in order to achieve the RDs cancellation. 
This is basically the main difference between the two approaches.
Although we have implemented the $\D$-regulator in all the results presented so far, however if we have used, for example, the regularization scheme of~\cite{Chay:2012mh} or the one in~\cite{Chiu:2012ir} then eq.~(\ref{eq:tmdnewdef}), with those regulators, would also give us a well-defined TMDPDF.

\section{Conclusions and Outlook}
\label{sec:conclusions}

We have provided a definition of TMDPDFs which is free from all rapidity divergences including the mixed terms that spoil the renormalization of such quantities.
The factorization theorem for DY $q_T$ dependent spectrum, which is the basis leading towards defining a ``TMDPDF'', is strongly believed to hold to all orders in perturbation and that the Glauber region is harmless. Since  full QCD quantities (like DY hadronic tensor) are  free from RDs, then given the analysis presented in this work, one can easily conclude that the RDs cancellation in the TMDPDF holds to all orders in perturbation theory.
This is important for phenomenological applications of TMDs (quarks in the case of DY or SIDIS or gluon TMDPDFs for LHC physics) since the anomalous dimensions and the $Q^2$-resummation kernel $D$~\cite{GarciaEchevarria:2011rb}, and thus the evolution of individual TMDs can now be properly determined~\cite{Echevarria:2012pw}.
Our results can be readily extended to define polarized and unpolarized quark and gluon TMDPDFs which could be considered as a generalization of the one introduced in~\cite{Echevarria:2012pw}, as well as TMD fragmentation functions.
The newly introduced definition of the TMDPDF reduces to the one proposed in~\cite{GarciaEchevarria:2011rb} when the two sectors are treated identically. 
We have also shown, by generalizing the arguments given in~\cite{Collins:2012uy}, how the new definition, which can be referred to as the ``modified EIS'' definition, is equivalent to the JCC one in the sense that both definitions manage to cancel rapidity divergences in bi-local quark fields separated along one light-cone direction and also in the transverse two-dimensional space.
Our definition can be readily used to carry out perturbative calculations beyond ${\cal {O}}(\alpha_s)$ (with any convenient regulator, the $\D$-regulator or any other one) and calculate explicitly, for example, the anomalous dimension of polarized (such as Sivers function) and unpolarized TMDs.

\section*{Acknowledgements}
This work is supported by the Spanish MECD, FPA2011-27853-CO2-02.
M.G.E. is supported by the PhD funding program of the Basque Country Government.
We would like to thank J. Collins and T. Rogers for useful discussions.
M.G.E. thanks LBNL for its hospitality while part of this work was completed.
A.I. would like to thank Andreas Sch\"afer for useful discussions.

\bibliography{references}


\end{document}